\documentclass[a4paper]{article}

\usepackage[english]{babel}
\usepackage[T1]{fontenc}

\usepackage[a4paper,top=3cm,bottom=2cm,left=3cm,right=3cm,marginparwidth=1.75cm]{geometry}

\usepackage{amsmath}
\usepackage{graphicx}
\usepackage[colorlinks=true, allcolors=blue]{hyperref}

\usepackage{textcomp}
\usepackage{authblk}

\title{Molybdenum Disulphide Nanoflakes Grown by Chemical Vapour Deposition on Graphite: Nucleation, Orientation, and Charge Transfer}

\author{Erik Pollmann}
\author{Juliana M. Morbec}
\author{Lukas Madau\ss}
\author{Lara Br\"{o}ckers}
\author{Peter Kratzer}
\author{Marika Schleberger}

\affil{Faculty of Physics and CENIDE, University of Duisburg-Essen, Lotharstra\ss e 1, Duisburg 47057, Germany}

\begin{document}
\maketitle

\begin{abstract}
Two-dimensional molybdenum disulphide on graphene grown by chemical vapour deposition is a promising van der Waals system for applications in optoelectronics and catalysis. To extend the fundamental understanding of growth and intrinsic properties of molybdenum disulphide on graphene, molybdenum disulphide on highly oriented pyrolytic graphite is a suitable model system. Here we show, experimentally and by density-functional-theory calculations, that molybdenum disulphide flakes grow in two orientations. One of the orientations is energetically preferred, the other one is rotated by 30 degree. Because of a high energy barrier confirmed by our calculations both orientations are stable at room temperature and their switching can only be forced by external stimuli, i.e. by a scanning tunneling microscope tip. Combined Kelvin probe microscopy and Raman spectroscopy measurements show that the flakes with a typical size of a few hundred nanometers are less doped than the often studied exfoliated molybdenum disulphide single layer. 
\end{abstract}

\section{Introduction}

Two-dimensional (2D) materials have become tremendously attractive research topics due to their extraordinary properties. Above all, graphene with its extraordinary thermal \cite{Balandin.2008,Kim.2010}, electrical \cite{Bolotin.2008} and mechanical \cite{Lee.2008} properties promises to revolutionize many applications. On the other hand, one of the biggest disadvantages for (opto)electronics is the zero band gap of graphene \cite{Novoselov.2004}. Other 2D materials such as molybdenum disulphide (MoS$_{2}$) have a band gap in the visible range \cite{Mak.2010,Kuc.2011}, but also a much lower conductivity \cite{Radisavljevic.2011}.

A combination of both materials, a so-called van der Waals heterostructure, can overcome their respective disadvantages by combining the advantages of both materials. The high conductivity of graphene together with the band gap of MoS$_{2}$ enables effective photoelectric devices \cite{Yu.2013,Zhang.2014,Chen.2016}. Graphene and MoS$_{2}$ based heterostructures are also potentially used in sensing \cite{Cho.2015,Pham.2019}, spintronics \cite{Dankert.2017}, or as memory devices \cite{Shinde.2014}. Such feasibility studies are performed with exfoliated 2D materials and recently also with 2D materials from chemical vapour deposition (CVD), which allows upscaling to large scale and flexible devices \cite{Yu.2014,Amani.2015}. To fabricate such heterostructure devices often stacking \cite{Dean.2010,CastellanosGomez.2014} and transfer techniques \cite{Li.2009,vanderZande.2013} are used. In contrast, direct growth of heterostructures provides a clean interface, but will at the same time have an effect on the physical properties.

In order to acquire a fundamental understanding of the growth and the intrinsic properties of MoS$_{2}$ on graphene, highly oriented pyrolytic graphite (HOPG) can be used as a model substrate. The HOPG surface has the same atomic surface structure as graphene and can be prepared much cleaner and with a lower defect density than the surface of CVD graphene, making HOPG ideal for studying the growth and the physical properties of MoS$_{2}$ on a sp$^{2}$ hybridized carbon surface.

When MoS$_{2}$ is grown by CVD, triangular islands -- which are called (nano)flakes in the following -- are typically formed. This particular shape is helpful in determining crystal orientations. Indeed, on graphite and on graphene preferred orientations of the MoS$_{2}$ flakes were found \cite{Liu.2016, Lu.2015, Koos.2016, Zhang.2017}. It is typically observed that MoS$_{2}$ flakes are rotated to each other by a multiple of 60 degrees, but in a few publications a rare orientation right inbetween has been reported. This has been experimentally shown by Liu et al. in the most comprehensive study \cite{Liu.2016}: Grazing-incidence wide-angle X-ray scattering proves that two orientations indeed occur for MoS$_{2}$ on epitaxial graphene (on silicon carbide) in different frequencies. The dominant orientation could be assigned to the case where the armchair direction of the MoS$_{2}$ flakes is aligned to the armchair direction of the underlying graphene layer. In the less frequent orientation, the zigzag direction of one lattice and the armchair direction of the other lattice correlate. On HOPG these orientation study have been done by comparing neighboring flakes in scanning probe microscope (SPM) images. Thereby it was found that MoS$_{2}$ grows mainly at the HOPG step edges \cite{Lu.2015, Koos.2016, Zhang.2017}. While experimental data on structure and growth is thus available, theoretical calculations elucidating the driving force behind the orientation behaviour are missing. Furthermore, studies regarding the charge transfer in MoS$_{2}$-based heterostructures are still rare. This is surprising as for many of the proposed electrical devices based on MoS$_{2}$-graphene heterostructures the electronic properties, such as carrier concentration (correlated with the Fermi level or doping), are a key factor.

In this paper, we address these issues and study ultrathin MoS$_{2}$ on HOPG in order quantify the carrier concentration in MoS$_{2}$. The paper is organized as follows: We first present structural data from our system and discuss an unexpected correlation between MoS$_{2}$ growth and the HOPG step height which we found by evaluating atomic force microscope (AFM) images. We then present an explanation for the experimentally observed MoS$_{2}$ flake orientation based on results from DFT calculations. Finally, we discuss the charge carrier concentration in MoS$_{2}$ which we derive by combining Raman and KPFM measurements. 

\section{Methods}
\subsection{Growth Process}
The samples were produced in a custom made process system for CVD of MoS$_{2}$ based on the system reported by Lee et al.~\cite{Lee.2012} and described in detail in our previous work \cite{Pollmann.2017}. Our process system consists of a quartz tube with two heating zones provided by a heating belt and a tube furnace (ThermConcept ROS 38/250/12). With this setup the two heating zones can be used with different temperatures to evaporate the molybendum source (MoO$_{3}$ powder, Alfa Aesar, 99.95\%) and the sulphur source (S powder, Sigma Aldrich, 99.98\%) separately. During the process inert Ar gas flows through the tube. Within the upstream heating zone of the heating belt a crucible containing sulphur powder is placed. In the second heating zone provided by the tube furnace a crucible with MoO$_{3}$ powder (upstream) and an HOPG crystal (downstream) is placed.

The recipe is as follows: 200~mg sulphur powder and 20~mg MoO$_{3}$ powder are used. A freshly cleaved HOPG crystal is placed 3~cm downstream from the molybdenum source. The heating rate of the tube furnace is 480~\textdegree C/min to reach the maximum temperature of 650~\textdegree C. When the furnace is started, the heating belt is activated with a temperature of about $\sim$70 \textdegree C to pre-anneal the sulphur powder and the Ar flow is set to 25~Ncm$^{3}$/min. Fifteen minutes before the furnace reaches the maximum temperature the belt is set to $\sim$150  \textdegree C and the Ar flow to 200~Ncm$^{3}$/min. While the heating zones stay at the maximum temperature a pressure of $\sim$6 mbar is measured in the quartz tube. Five minutes after the furnace has reached the maximum temperature the furnace and the belt are switched off to cool the system down. In this procedure MoS$_{2}$ single layers grow on the face-down side in the microcavity between HOPG crystal and crucible.

\subsection{Characterization}
The samples are characterized with a SPM operating in a RHK UHV 7500 system with a base pressure of about 4~*~10$^{-10}$~mbar. This ensures that the influence of contaminations during measurements is mimimized. The SPM is controlled by the fully integrated system for digital signal processing RHK R9. The system is used to perform non-contact atomic force microscopy (NC-AFM) and Kelvin probe force microscopy (KPFM) simultaneously, and scanning tunneling microscopy (STM). For NC-AFM and KPFM commercial tips are used (Nanoworld NCH). The tips for performing STM are etched with the lamella method, for which a tungsten wire and a 5~molar NaOH solution are used. The Raman spectra are recorded using a Renishaw InVia Raman microscope at the Interdisciplinary Center for Analytics on the Nanoscale (ICAN, core facility funded by the German Research Foundation, DFG). The laser wavelength is 532~nm and the spot diameter is 1~\textmu m. The laser power was kept at 200~\textmu W to avoid damage. The Raman microscope was calibrated on a silicon sample with a Si Raman mode at 520.78~cm$^{-1}$.

\section{Results and Discussion}

\subsection{Growth morphology and nucleation}

\begin{figure} [thb]
\centering
\includegraphics[width=1.0\textwidth]{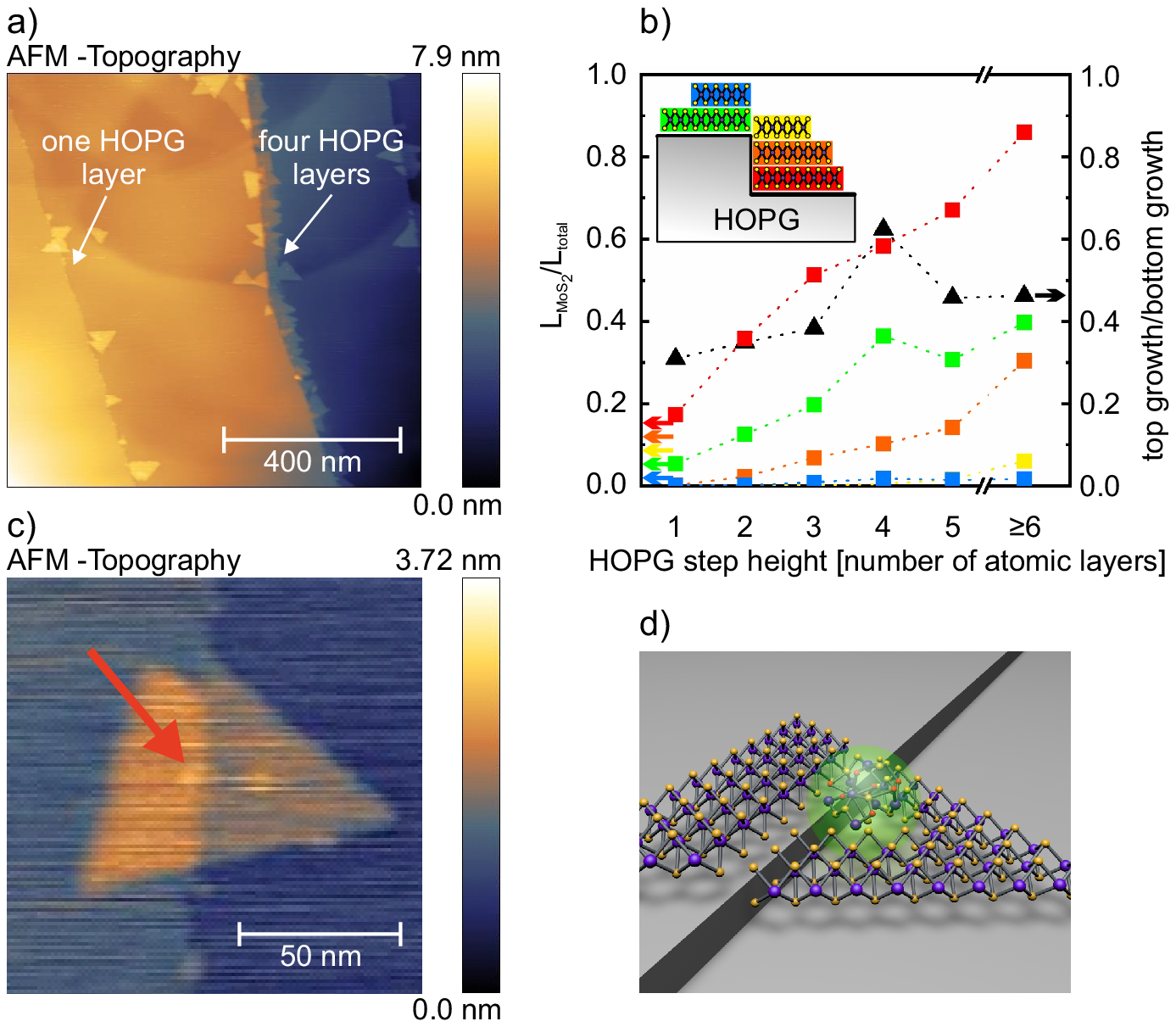}
\caption{a) Layered MoS$_{2}$ growth on HOPG substrate observed in a typical AFM image of the surface after the growth procedure. The heights of the two HOPG steps correspond to one atomic layer and four atomic layers, respectively. Triangular shaped MoS$_{2}$ layers are most frequently found at these HOPG steps, preferably at higher steps. b) Diagram showing the step height dependency of the growth. The coloured lines show the ratio of the measured step edge length decorated with MoS$_{2}$ to the total measured length. The different colours indicate different species as shown in the inset. The black triangles show the ratio between the growth on the upper and lower  ledge. c) Zoomed AFM image of a MoS$_{2}$ flake with a nanoparticle (indicated by the red arrow) located at the step edge. d) Illustration of the arrangement of HOPG substrate, nanoparticle and the MoS$_ {2}$ layers at the step.}
\label{Fig:Growth}
\end{figure}

Our investigations have shown that MoS$_{2}$ does not grow as a continuous film on graphene, but rather in the form of islands. We start our study by discussing this peculiar growth morphology. In Fig. \ref{Fig:Growth} (a) and Fig.~\ref{Fig:Orientation}~(a) AFM images of a sample after the growth procedure are shown. The HOPG surface with its typical edges of one to several atomic sp$^{2}$ hybridized carbon layers is still intact. Furthermore, triangular shaped MoS$_{2}$ nanoflakes with edge lengths of up to 100~nm can be observed. With the growth conditions discussed above we make two observations: (i) most of the MoS$_{2}$ nanoflakes are located at the HOPG edges and (ii) are found in two orientations.

At first we discuss the preferred occurrence of the MoS$_{2}$ flakes at HOPG edges, which has not been previously analysed any further to the best of our knowledge, although it is obviously crucial for growth. The AFM image in Fig.~\ref{Fig:Growth} (a) already gives first evidence for a correlation between the step height and the MoS$_{2}$ flake height. The HOPG step edge with a height corresponding to four layers in the right half of the image is decorated by a continuous sequence of multilayer MoS$_{2}$ islands, while only a few monolayer MoS$_{2}$ triangles are located at the mono-atomic HOPG step edge in the left half of the image. Moreover, we observe island growth both at the upper and the lower step edge, but with more material accumulated at the lower step edge. In order to quantify this observation we analysed different areas on our sample by correlating (i) the HOPG step edge height with the MoS$_{2}$ flake height and (ii) the ratio between the HOPG step length decorated with MoS$_{2}$ and the total measured edge length $L_{MoS_{2}}/L_{total}$. In total, step edges of more than 300~\textmu m in length were evaluated. The result is shown in Fig.~\ref{Fig:Growth} (b), where $L_{MoS_{2}}/L_{total}$ is plotted versus the HOPG step height in atomic layer numbers. Different colours indicate whether the MoS$_{2}$ flakes had grown on the lower (red, orange, yellow) or on the upper terrace (green, blue) of the HOPG step edge (see illustration in the inset of Fig.~\ref{Fig:Growth} (b)). The dashed lines are included to guide the eye. The results show a clear correlation between the HOPG step edge height and the MoS$_{2}$ flake height, and the ratio $L_{MoS_{2}}/L_{total}$, respectively, and thus the total amount of grown MoS$_{2}$.

The observation that the MoS$_{2}$ flake height differs on the upper and the lower side of the HOPG step edge indicates that the step edges constitute barriers for materials transport, a phenomenon known as the  Ehrlich-Schw\"{o}bel effect \cite{Ehrlich.1966,Schwoebel.1966,Schwoebel.1969}. 
It remains to be discussed to what extent the results for  the ratio $L_{MoS_{2}}/L_{total}$ could be affected by overgrowth, i.e., an MoS$_{2}$ island could nucleate on the upper terrace and then extend to grow on top of a multilayer island of the lower terrace.  Let us look at the ratio between layers grown on upper and lower terraces (black data points in Fig.~\ref{Fig:Growth} (b) ). This ratio is almost independent of the HOPG step height (discarding the peak at four-layers HOPG steps, see  Supp. Mat. for justification). If overgrowth of islands nucleated on the upper terrace was to play a major role, one would expect a clear variation, since it would become more and more difficult to overgrow the HOPG step with increasing step height. Thus, at least for HOPG steps higher than one layer, we believe that island growth at the upper and lower ledges follows  separate nucleation events.

In order to understand the observed peculiarities, we look at the possible nucleation mechanisms next. For the present case, homogeneous nucleation by  random accumulations of material \cite{Zhu.2017}, or heterogeneous nucleation by nanoparticles (NP) \cite{Cain.2016, Zhu.2017} need to be taken into consideration. It is safe to assume that these mechanism could exist simultaneously, but that their respective contribution may differ under different conditions. Other suggested mechanisms, such as so-called seeding promoters (aromatic molecules acting as artificial nucleation centers) \cite{Lee.2012, Ling.2014}, or artificially induced defects by ion irradiation \cite{Pollmann.2018} are not applicable to our present study.

In fact, several pieces of evidence point to the role of nanoparticles as nucleation centers. For instance, Cain {\em et al.}~reported MoO$_{3-x}$S$_{y}$ nanoparticles present during the early stages of MoS$_{2}$ growth, when the sulphur concentration is not high enough \cite{Cain.2016}. These nanoparticles diffuse on the surface and it is plausible that they get trapped at the HOPG step edges. Larger nanoparticles will more likely be trapped at higher step edges leading to a higher concentration there. As the layer growth starts, MoS$_{2}$ flakes would preferentially grow at these nanoparticles. Since a nanoparticle may well be higher than the edge itself (see illustration in Fig. \ref{Fig:Growth} (d)), it is able to promote growth of  MoS$_{2}$ flakes on the lower as well as on the upper terrace, in agreement with our data (see Fig. \ref{Fig:Growth} (b)).  This hypothesis agrees also very well with our observation that there is never growth on an upper terrace without growth on the adjoining lower terrace. In addition, the nanoparticle hypothesis is backed by our experimental observations: Figure \ref{Fig:Growth} (c) shows a zoomed AFM image of a non-perfect triangular MoS$_{2}$ flake. It consist of a trapezoidal part at the upper terrace and a triangular part at the lower terrace separated by the step edge. Where both flakes are touching at the position of the edge, a bright spot is seen (red arrow). A few other similar spots could be seen at the surface, but most of them are located at HOPG edges. We propose that these bright spots correspond to the nanoparticles  acting as nucleation centers.  During later stages of the growth, the nanoparticles could be consumed by the growth of MoS$_{2}$ \cite{Zhu.2017}. This explains our finding that some flakes do not have any bright spots in their centers, like the ones seen in Fig.~\ref{Fig:Growth} (b).

\subsection{Orientation}
\label{Chapter:Orientation}

\begin{figure} [thb]
\centering
\includegraphics[width=1.0\textwidth]{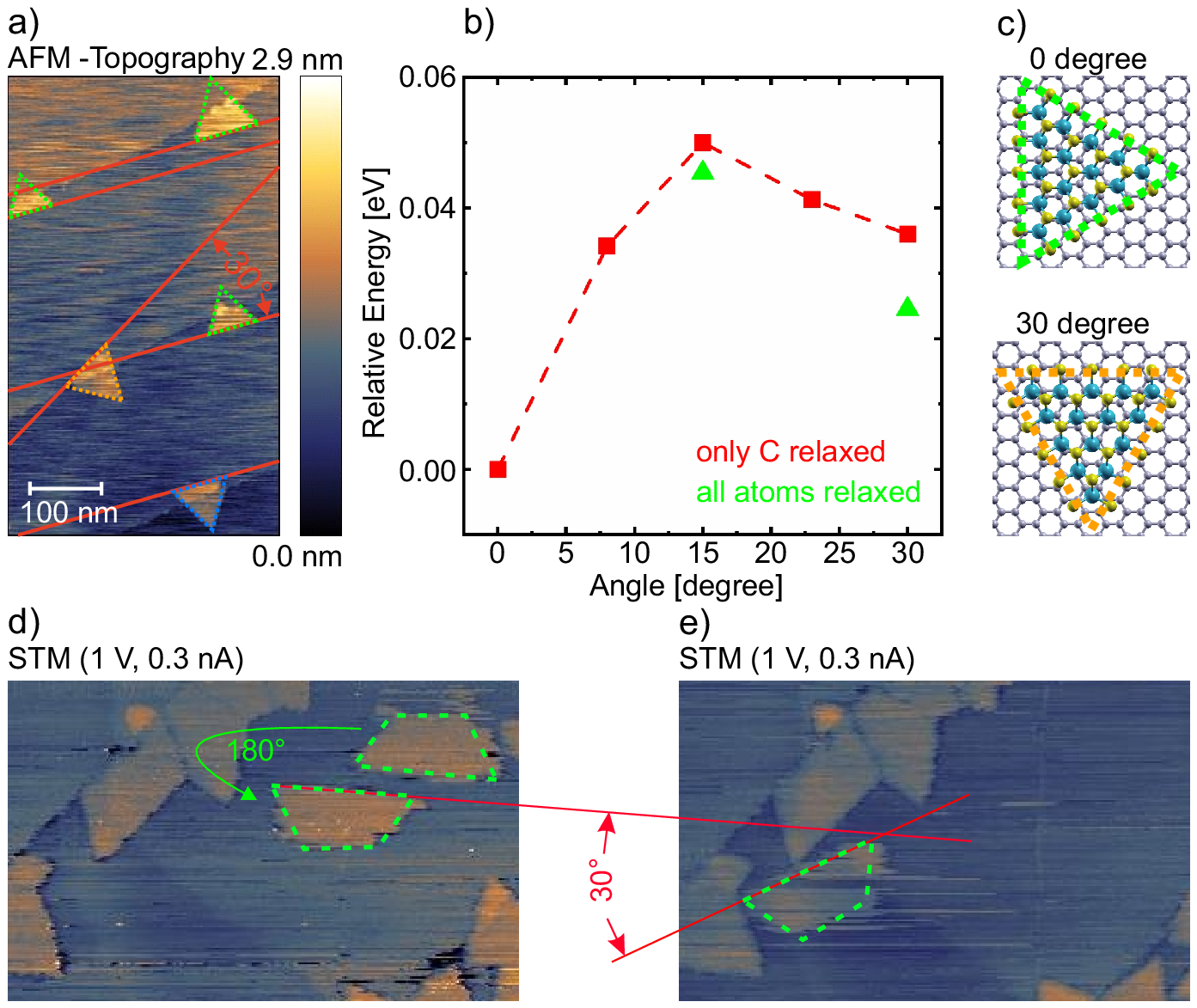}
\caption{a) Image taken by AFM showing MoS$_{2}$ triangles on HOPG surface. The flakes are found in two main orientations, rotated by 30 degree with respect to each other. (b) Relative total energies, computed with DFT calculations and considering different geometry relaxation conditions, for different angles between the armchair direction of the MoS$_{2}$ flake and the armchair direction of the graphene sheet. (c) Schematic view of the 0 degree (armchair direction of the MoS$_{2}$ flake aligned with armchair direction of the graphene sheet) and 30 degree (armchair direction of the MoS$_{2}$ flake aligned with zigzag direction of the graphene sheet) orientations. (d) and (e) STM images (sample bias 1~V, tunneling current 0.3~nA) showing the MoS$_{2}$ crystallites being moved by the STM tip; rotation MoS$_{2}$ flake in (d) and sliding MoS$_{2}$ flake in (e).}
\label{Fig:Orientation}
\end{figure}

Next, we turn to the later stages of growth and look at fully developed MoS$_{2}$ flakes. During growth, an epitaxial relationship concerning the orientation of the flakes is built up. To analyse this orientation, we focus on large MoS$_{2}$ flakes located on HOPG terraces or single-layer steps. As marked in Fig.~\ref{Fig:Orientation} (a) most of the MoS$_{2}$ flakes are pointing in the same direction (green triangles). A significant fraction is however rotated with respect to this main orientation by either 30 degrees (orange) or 60 degrees (blue). It is noteworthy that under similar conditions MoS$_{2}$ grows on (amorphous) SiO$_{2}$ with edge length of several microns and without any preferential orientations \cite{Pollmann.2017}. Obviously, it is the crystallinity of the HOPG substrate which gives rise to the preferred orientations of the MoS$_{2}$ flakes. Because of the rotational symmetry of the hexagonal graphite lattice, the orientation of the green and blue triangles corresponds to equivalent crystallographic directions. By comparing the edges of neighbouring flakes, it can be determined that the two orientations have different frequencies of occurrence (see Supp. Mat. for more AFM images). Accordingly, in Fig.~\ref{Fig:Orientation}~(a) the green/blue orientation is the preferred one and the orange the less preferred one. 

It is instructive to compare our findings to observations in other van-der-Waals-bonded systems. For graphene flakes on a graphene surface, which exhibits the same sp$^{2}$ hybridized carbon surface as the HOPG surface used here, it has been also experimentally shown that the flakes have a preferred orientation \cite{Feng.2013}. However, these flakes were found only in orientations of multiples of 60~degree, because minima occur in the potential landscape at these orientations \cite{Shibuta.2011}.

To explore the potential energy surface for the MoS$_{2}$-HOPG system, DFT calculations were performed (for details see Supp. Mat). To make the computations feasible, a rather small flake size (15~Mo + 50~S atoms, $\sim$1~nm edge length) has been used, and the substrate is described by a single layer of graphene instead of HOPG. Triangular flakes with Mo-edges, terminated by 100\% sulphur, were cut from a two-dimensional MoS$_{2}$ layer, as these edges are known to be energetically stable for the sulphur-rich environment in STM investigations of MBE grown MoS$_{2}$ \cite{Helveg.2000, Schweiger.2002}. Several registries between the MoS$_{2}$ flakes and the graphene have been used as initial geometries, and the atomic positions were optimized using the forces obtained from the DFT calculations. The geometries shown in Fig.~\ref{Fig:Orientation} (c) are the most stable configurations found in this way. Starting from these initial configurations, the flakes are rigidly rotated, allowing for optimization of the C-atom positions for each orientation. 

For symmetry reasons, rotation angles in the range of 0 to 30~degree suffice, where 0~degree corresponds to orientational alignment, i.e. the armchair direction of both lattices, but also the zigzag directions (e.g. the edges of the MoS$_{2}$ flakes)  coincide,  whereas 30~degree corresponds to a match of armchair-zigzag between both lattices. The results are shown in Fig.~\ref{Fig:Orientation}~(b), where the energy relative to the energy for the 0~degree configuration is plotted as function of the rotation angle. The figure clearly shows that the 0~degree (armchair-armchair) orientation is the energetically most favorable configuration, but there is also a local minimum for the 30~degree (armchaior-zigzag) orientation, which agrees with the experimental data of the present study and of Liu {\em et al.} \cite{Liu.2016}. The bending of the graphene sheet and the downward movement of the C atoms beneath the MoS$_{2}$ flake play a significant role for the stability of the flake in both orientations. If the C atoms were {\em not}  allowed to re-adjust, the increase in energy when the flake rotates from 0 to 30~degree would be considerably larger than for relaxed C atoms (cf. Supp. Mat. for more details). We learn from the calculations that the rotation goes along with an increased adsorption height of the MoS$_{2}$ flake, and hence a reduced van der Waals binding energy. In large MoS$_{2}$ flakes the formation of a commensurate superstructure is to be expected, as has been reported for MoSe$_2$ on few-layer graphene \cite{Dau.2018}. The potential energy surface has periodically repeated minima at spots where one sulphur atom sits right above a hollow site of the graphene lattice, while the neighbouring sulphur atoms are still close to neighbouring hollow sites. For the 30 degree-rotated configuration, such a favorable match is no longer possible; even in the optimized structure some fraction of the S atoms is forced to sit on top of C atoms. The steric repulsion between these atoms enforces the enhanced adsorption height in case of the rotated flake. The calculation for the condition of relaxation of all atoms (green triangle) only slightly differs from the calculation with the rigid MoS$_{2}$ flake; thus the elastic response of the graphene layer to the orientation of the flake is the most crucial factor for the obtained potential landscape with local minimum at 30~degree.

The fact that the MoS$_{2}$ flakes in our AFM images (Fig.~\ref{Fig:Orientation} (a)) are also found in the less probable armchair-zigzag orientation indicates that this orientation is at least metastable at room temperature. Not even the additional force from the AFM tip is sufficient to overcome the energy barrier for a rotation of the MoS$_{2}$ flakes, because a switching of flakes between two orientations has never been observed during our NC-AFM measurements. Indeed, already the very small flake modeled in Fig.~\ref{Fig:Orientation} (c) shows a barrier at 15 degree towards rotation of 14~meV (for all atoms allowed to relax 21~meV). For realistic flakes this energy barrier will be much higher, and thus clearly exceed the thermal energy of $k_B T = 25$~meV at room temperature. Note that it is difficult to estimate the barrier for rotating a large flake since edge effects in the calculations for the small flakes may lead to an overestimation of the barrier when scaled to larger flakes.

Our finding that differently oriented flakes correspond to different binding energies can be further corroborated by STM measurments which can be used to force sliding and rotating MoS$_{2}$ flakes across the HOPG surface (in contrast to the AFM) during imaging. To force reorientations (flips) we apply a technique reported by Feng {\em et al.}, who could successfully manipulate graphene flakes on graphene by STM \cite{Feng.2013}, and by B\"{u}ch {\em et al.} for WS$_{2}$ flakes on graphene \cite{Buch.2018}. The mechanism Feng et al. proposed for the reorientation is a vertical displacement of the flakes by the interaction with the STM tip as reported by Wong et al. \cite{Wong.2009}.

Figure \ref{Fig:Orientation} (d) and (e) show subsequently captured STM images of our MoS$_{2}$-HOPG system recorded at 1~V sample bias and 0.3~nA tunneling current of the same surface area. The flake marked in green is not fully triangular, and we can make use of its characteristic form for the following analysis. The images were recorded beginning from the upper left corner downwards and it shows that this particular flake rotates during the first image acquisition by 180 degree as depicted by the green arrow. The second image shows that the flake has rotated by a further 30~degree. In addition the flake obviously slides over the surface during the scans. As a telltale sign one can clearly see the bright stripes along the fast scan direction to the right of the marked area in Fig. \ref{Fig:Orientation} (e). In all cases only rotations of multiples of 30 degree were observed, which confirms our calculation of the two preferred orientations. Translations were often observed in combination with rotations. Especially flakes found in the 30~degree orientation are susceptible to translation. We observe much larger sliding distances as in the case of graphene \cite{Feng.2013}, i.e. several 100~nm. That the forced flake movement and 30~degree orientation is closely interrelated confirms that the MoS$_{2}$ flakes in this configuration has a binding energy to the underlying substrate which is weak enough to allow sliding of the flakes in this metastable orientation.

\begin{figure}[htb]
\centering
\includegraphics[width=1.0\textwidth]{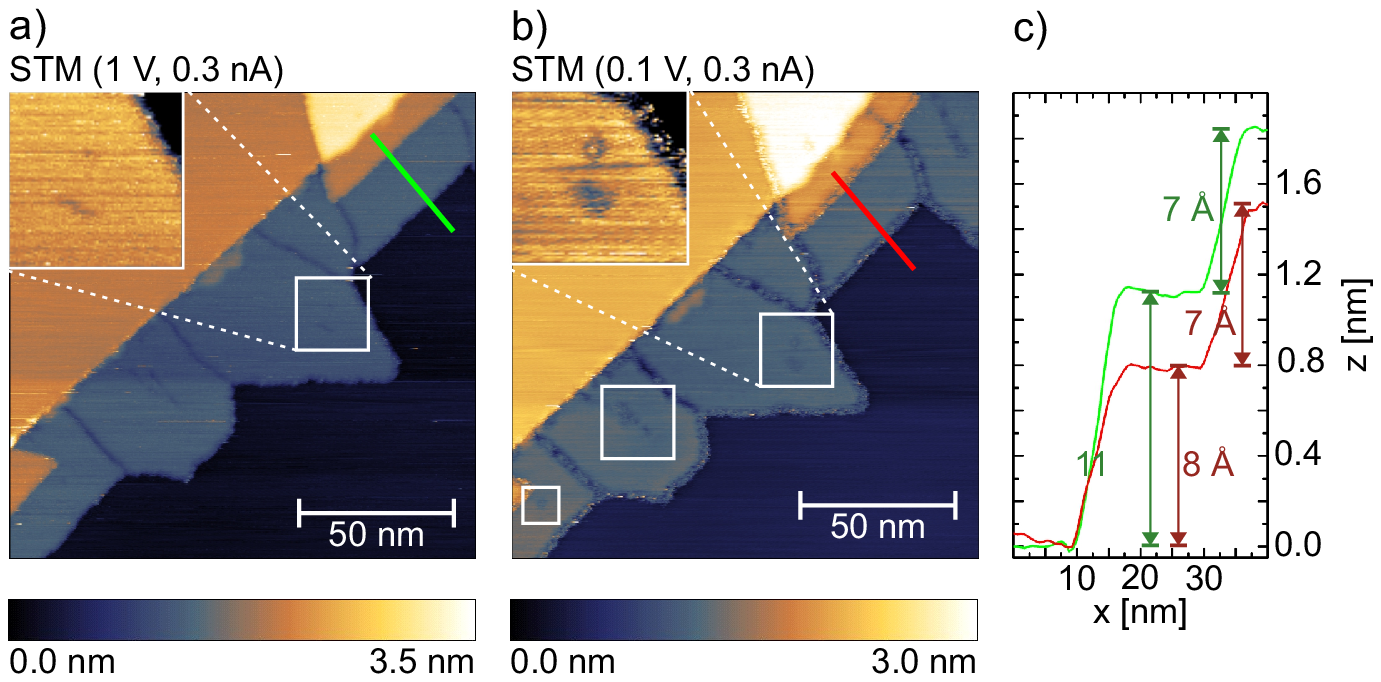}
\caption{Parameters for STM may be tuned to make defects more apparent: 
a) Topographic STM mode (1~V, 0.3~nA), which looks similar to AFM pictures. b) Mode which reveals defects and grain boundaries (0.1~V, 0.3~nA). c) Extracted profile lines from the green and red line in a) and b). With both sample biases, the second jump (by 7 {\AA}) corresponds to one MoS$_{2}$ layer step. The first jump is higher and voltage-dependent. We interpret this as an indication for the vertical displacement of the monolayer MoS$_{2}$ flake by the STM tip which seems to be possible because the van der Waals interaction of MoS$_{2}$ with HOPG is weaker compared to the interaction between two MoS$_{2}$ layers.}
\label{Fig:Defects}
\end{figure}

The ability of the STM tip to lift up the flake is corroborated by the fact that with STM -- in contrast to NC-AFM -- we frequently find single-layer MoS$_{2}$ step edges which appear higher than the nominal thickness of MoS$_{2}$, see Fig. \ref{Fig:Defects} (a)-(c) (in addition, the different sample voltages make the STM more sensitive to either the topography Fig.~\ref{Fig:Defects}~(a) or to the antiphase domain boundaries and defects (b), which can under certain conditions also be induced by the STM tip itself (see Supp. Mat)). This observation may be seen as an indication that the interaction of the STM tip with the MoS$_{2}$ flake is comparable in strength to the interaction between MoS$_{2}$ and graphite. Once the flake has been relocated by the tip into the local minimum (see Fig. \ref{Fig:Orientation} (c)), the global minimum can be reached at room temperature only after another tip-flake interaction to overcome the energy barrier.

\subsection{Carrier concentration in MoS$_{2}$}

\begin{figure}[htb]
\centering
\includegraphics[width=1.0\textwidth]{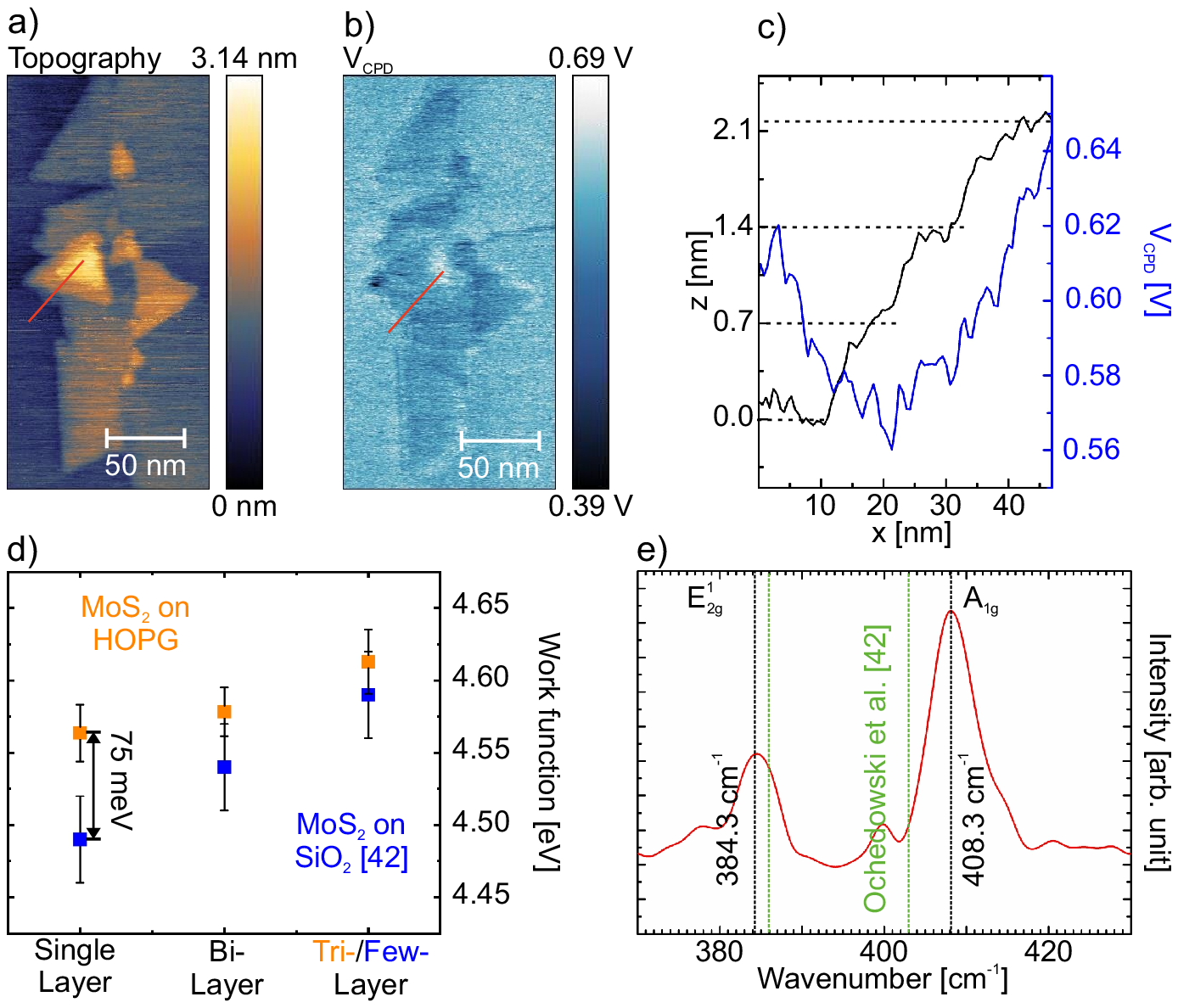}
\caption{Indications for electronic doping can be either obtained by (i) spatially resolved KPFM-measurement or by (ii) averaging over several flakes with Raman spectra. a) Topographic AFM image and b) $V_{\rm CPD}$ image obtained by KPFM of MoS$_{2}$ flakes on HOPG. Bi- and trilayer areas are clearly seen. c) Profiles along the red lines in a) and b) reveal the steps of MoS$_{2}$ and a layer number dependency of the $V_{\rm CPD}$ value, which is plotted in d) with respect to exfoliated MoS$_{2}$ on SiO$_{2}$ \cite{Ochedowski.2014}. e) The MoS$_{2}$ Raman modes are located at 384.3 cm$^{-1}$ and 408.3 cm$^{-1}$, respectively. The green dashed lines indicates the positions for single layer MoS$_{2}$ on SiO$_{2}$ reported by Ochedowski et al. \cite{Ochedowski.2014}.}
\label{Fig:Doping}
\end{figure}

In the final section, we want to discuss the electronic properties of our samples. To this end we combined Raman spectroscopy with a laser spot size of 1~\textmu m as an averaging technique with respect to the flake size with KPFM measurements yielding a high spatial resolution. The KPFM measurements are performed on our samples in UHV to exclude any influence of the environment \cite{Ochedowski.2014}. Fig. \ref{Fig:Doping} (b) shows the contact potential difference ($V_{\rm CPD}$) image of single layer, bilayer, and trilayer MoS$_{2}$ flakes, which are clearly seen in the topographic image in Fig. \ref{Fig:Doping} (a). In the $V_{\rm CPD}$ image a material contrast between MoS$_{2}$ and HOPG is observable. The single layer and bilayer areas are almost indistinguishable, while the trilayer has an increased $V_{\rm CPD}$ value with respect to the HOPG surface. The linescans in Fig. \ref{Fig:Doping} (c) obtained along the red line in the topography and $V_{\rm CPD}$ images show this more clearly. The dotted lines indicate the MoS$_{2}$ layer height of about 0.7~nm, which confirms the number of MoS$_{2}$ layers in the image.

Because the work function of HOPG is very well known and HOPG is inert, we can easily calibrate our data against this value. With $\phi_{\rm HOPG}$ = 4.6~eV \cite{Takahashi.1985} we obtain the following work function for single layer $\phi_{\rm SL-MoS_2} \approx$ 4.564 $\pm$ 0.02 eV, for bilayer $\phi_{\rm BL-MoS_2} \approx$ 4.578 $\pm$ 0.02 eV, and for trilayer MoS$_{2}$ $\phi_{\rm TL-MoS_2} \approx$ 4.613 $\pm$ 0.02 eV. These values are in good agreement with the study of Precner et al. on MoS$_{2}$ on HOPG \cite{Precner.2018}. In order to classify the obtained values in terms of the relative charge transfer our results are compared with the work function of exfoliated MoS$_{2}$ flakes on SiO$_{2}$, which have been characterized under similar conditions with the same KPFM technique \cite{Ochedowski.2014}, see Fig. \ref{Fig:Doping} (d). In both cases, MoS$_{2}$ on HOPG and MoS$_{2}$ on SiO$_{2}$, the work function increases with the number of layers. However, the work function for single layer MoS$_{2}$ on HOPG is shifted towards higher values by approximately 75 meV. The Fermi level $E_{F}$ and, by implication, the work function too, can be related to the electron density $n$. Thus, we find that in the case of MoS$_{2}$ on HOPG electrons are effectively depleted in the MoS$_{2}$ layer in comparison to MoS$_{2}$ on SiO$_{2}$.

In order to quantify the KPFM data the relation for the Fermi level of a non-degenerate semiconductor to its electron density is used: $E_{F}~=~E_{Fi}~+~k_{B}T \, \ln(n/n_{i})$, with $E_{Fi}$ the intrinsic Fermi level, $T$ the temperature and $k_B$ the Boltzmann constant. Thus, the charge carrier density can only be determined with respect to the intrinsic Fermi level $E_{Fi}$ and the intrinsic electron density $n_{i}$. Although $E_{Fi}$ and $n_{i}$ are unknown, the ratio $n_{\rm HOPG}/n_{\rm SiO_2}$ as the doping concentration ratio between MoS$_{2}$ supported by HOPG and by SiO$_{2}$ can be determined. For single layer MoS$_{2}$ with a Fermi level change of $\Delta E_{F} \approx$ 75 meV we obtain a ratio of $n_{\rm HOPG}/n_{\rm SiO_2} \approx$ 0.055. This means, that the electron density in MoS$_{2}$ on SiO$_{2}$ due to doping by charge transfer is one to two orders of magnitude higher than in MoS$_{2}$ on HOPG.

Without $E_{Fi}$ or $n_{i}$ the relative order of magnitude of the doping of both systems can be determined, but the absolute number of the order of magnitude remains unknown. Therefore, Raman measurements are performed. The spectra shown in Fig. \ref{Fig:Doping} (e) exhibit the characteristic peaks for MoS$_{2}$: the E$^{1}_{2g}$ and the A$_{1g}$ mode are found at 384.3 cm$^{-1}$ and 408.3 cm$^{-1}$, respectively. Because several AFM images (e.g. Fig. \ref{Fig:Growth} (a)) and also the diagram in Fig. \ref{Fig:Growth} (b) show that the most part of the MoS$_{2}$ film has grown as single layer, we take the average Raman signal from several such small flakes and again compare this data with data taken from larger single layer MoS$_{2}$ flakes exfoliated on SiO$_{2}$. There are several Raman studies on MoS$_{2}$ on SiO$_{2}$, including that by Lee et al. \cite{Lee.2010} and own previous studies \cite{Ochedowski.2014, Pollmann.2017}. Taking into account and adjusting different offsets during calibration, the E$^{1}_{2g}$ and the A$_{1g}$ mode for MoS$_{2}$ on SiO$_{2}$ can be found in the range of 384.6 - 386.5~cm$^{-2}$ and 403 - 405.4~cm$^{-2}$, respectively. For quantitative analysis, the Raman data of Ochedowksi {\em et al.} are used for comparison, as this was already used as a comparative study for the KPFM data \cite{Ochedowski.2014}. These reference peak positions at 386.1~cm$^{-2}$ and 403.0~cm$^{-2}$ are plotted with a green dotted line in the diagram of Fig. \ref{Fig:Doping} (e). The slightly lower A$_{1g}$ mode in contrast to other studies may originate from pre-processing of the SiO$_{2}$ substrate in this particular study. The peak difference for MoS$_{2}$ on HOPG is obviously larger than for MoS$_{2}$ on SiO$_{2}$. A Raman peak shift of the MoS$_{2}$ modes can have different origins. Apart from the number of MoS$_{2}$ layers \cite{Lee.2010}, e.g. strain \cite{Rice.2013, Pollmann.2017} and electronic doping \cite{Chakraborty.2012} are known to affect the peak positions in different ways. The latter affects almost exclusively the A$_{1g}$ mode, while strain is more influential on the E$^{1}_{2g}$ mode, and the layer number affects both Raman modes. In qualitative terms, this means that MoS$_{2}$ on HOPG is more stretched (or less compressed) and undergoes electron depletion with respect to MoS$_{2}$ on SiO$_{2}$.

The linear dependency of the Raman mode shift on both, strain and electron accumulation, experimentally and theoretically determined and reported by Rice {\em et al.}~\cite{Rice.2013} and Chakraborty {\em et al.}~\cite{Chakraborty.2012}, respectively, allows a rapid quantitative analysis of the data. The E$^{1}_{2g}$ mode shifts by $-2.1$~cm$^{-2}$ per \% of strain and by 0.6~cm$^{-2}$ per electron accumulation of $1.8 \times 10^{13}$~cm$^{-2}$, the A$_{1g}$ mode shifts by $-0.4$~cm$^{-2}$ per \% of strain and by 4~cm$^{-2}$ per electron accumulation of $1.8 \times 10^{13}$~cm$^{-2}$. Based on their data, we can estimate a relative strain of $\sim$1.3 \% and the change of electron density for the MoS$_{2}$-HOPG system with respect to the MoS$_{2}$-SiO$_{2}$ system to be $\Delta n~=~n_{\rm HOPG}~-~n_{\rm SiO_2}~\approx -2.6 \times 10^{13}$/cm$^{2}$. As our Raman microscope averages over several flakes within one measurement and a Raman shift could be due to contributions from few-layer MoS$_{2}$, this value may not be exact but will be correct within the order of magnitude. Combined with the estimation from the KPFM data we find the doping of MoS$_{2}$ on SiO$_{2}$ to be on the order of 10$^{13}$/cm$^{2}$, while the doping of MoS$_{2}$ on HOPG is on the order of 10$^{12}$/cm$^{2}$.

We propose that this difference is due to the following reasons: The high concentration of impurities in SiO$_{2}$ and the intercalated water film, which is typical for exfoliated flakes, leads to a different charge-transfer into the MoS$_{2}$ from the SiO$_{2}$ than from the HOPG. In contrast, those doping mechanisms are absent on the clean HOPG surface. The MoS$_{2}$ layers were grown on the HOPG substrate under such high temperatures that no water film is intercalated, therefore the layer thickness of the first MoS$_{2}$ layer corresponds to the expected value of about 0.7 nm, see Fig.~\ref{Fig:Doping} (c). Thus, MoS$_{2}$ on a quasi pristine and conductive HOPG surface is much closer to its intrinsic electronic state. Note that we do not know the intrinsic Fermi level nor the intrinsic charge carrier concentration. A much stronger hole doped MoS$_{2}$ on HOPG with respect to MoS$_{2}$ on SiO$_{2}$ would also be in agreement with our data.

Another possible reason for the difference in the doping level is that CVD MoS$_{2}$ may have a higher defect density than exfoliated MoS$_{2}$. The formation of sulfur vacancies is much more likely than molybdenum vacancies in CVD and exfoliated MoS$_{2}$ \cite{Hong.2015, Komsa.2012, Madau.2018}. These sulfur vacancies have no influence on the doping sensitive Raman mode of single layer MoS$_{2}$ \cite{Parkin.2016} and as a consequence would escape our analysis here. Furthermore, various photoluminescence studies have shown that intercalated water on the SiO$_{2}$ substrate has a strong electron doping influence on single layer MoS$_{2}$ \cite{Scheuschner.2014, Buscema.2014, Sercombe.2013} rendering the different substrate and environment of the MoS$_{2}$ to be the most likely reason of the different carrier concentration.

\section{Summary}

Van der Waals heterostructures with their peculiar interface properties offer new opportunities for novel device architectures. Their fabrication poses however still a challenge as stacking and transfer techniques come along with rather ill-defined inter- and surfaces. Direct growth of such heterostructures provides an excellent alternative in this respect but optimum growth parameters as well as properties remain to be investigated. In this work, two-dimensional MoS$_{2}$ has been grown directly on HOPG by CVD and studied by complementary methods. A self-seeding nucleation process was identified to be at the origin of the growth mechanism. We found two stable flake orientations which could be assigned by our DFT results to the more frequent armchair-armchair configuration and the less frequent armchair-zigzag configuration. A reorientation between the two configurations takes place only if an external stimulus is applied. In comparison to MoS$_{2}$ grown on SiO$_{2}$, MoS$_{2}$ grown on HOPG exhibits a much lower doping level due to charge transfer. The charge carrier concentration of MoS$_{2}$ on HOPG is presumably close to its intrinsic value. However, for a conclusive determination a quantitative analysis of suspended MoS$_{2}$ samples under UHV conditions via KPFM would be required -- an experimental challenge which has yet to be met .

\section*{Acknowledgements}
J.M.M. and P.K. gratefully acknowledge the computing time granted by the Center for Computational Sciences and Simulation (CCSS) of the University of Duisburg-Essen and provided on the supercomputer magnitUDE (DFG Grant No. INST 20876/209-1 FUGG and INST 20876/243-1 FUGG) at the Zentrum f{\"u}r Informations- und Mediendienste (ZIM). M.S. and L.B. acknowledge support from the DFG by funding SCHL 384/20-1 (project number 406129719) and project C5 within the SFB1242 \glqq Non-Equilibrium Dynamics of Condensed Matter in the Time Domain\grqq \ (project number 278162697). Raman measurements were carried out by L.M. at the Interdisciplinary Center for Analytics on the Nanoscale (ICAN), a core facility funded by the German Research Foundation (DFG, reference RI\_00313).

\bibliographystyle{unsrt}
\bibliography{Literatur}

\newpage

\end{document}